\documentstyle[12pt]{article}
\begin{document}
\renewcommand{\thefootnote}{\fnsymbol{footnote}}
\begin{flushright}
TIFR/TH/97-27\\
\end{flushright}
\vspace*{3cm}
\begin{center}
{\Large \bf \boldmath $h_c$ production at the Tevatron
\footnote{Presented at the XXXII Rencontres de Moriond, 
``QCD and High Energy Hadronic Interactions'', March 22-29, 1997, 
Les Arcs, France.}} \\
\vskip 32pt
{\bf K. Sridhar\footnote{sridhar@theory.tifr.res.in}}\\
{\it Theory Group, Tata Institute of Fundamental Research, \\ 
Homi Bhabha Road, Bombay 400 005, India.}

\vspace{100pt}
{\bf ABSTRACT}
\end{center}
\vspace{12pt}

Non-Relativistic QCD (NRQCD) has provided a consistent framework
for the physics of quarkonia. The colour-octet Fock
components predicted by NRQCD have important implications for
the phenomenology of quarkonium production in experiments. We
have considered the consequences of the colour-octet channels
for the production of the ${}^1P_1$ charmonium state, $h_c$, at 
large-$p_T$ at the Tevatron. The colour-octet contributions to
this state are found to be dominant and give a reasonably large
rate for the production of $h_c$. This should make it feasible to 
look for this resonance in the $J/\psi +\pi$ decay channel. The
observation of $h_c$ is interesting in its own right, since its
existence has yet to be experimentally confirmed. Moreover, this
rate is a $prediction$ of NRQCD, and the observation of
the $h_c$ at the Tevatron can, therefore, be used as a test 
of NRQCD.

\vfill
\clearpage
\pagestyle{empty}
$J/\psi$ production has traditionally been described in
terms of the colour-singlet model \cite{berjon, br}, which, 
in spite of its reasonable success with its predictions at lower
energies, fails completely in explaining the data on 
large-$p_T$ $J/\psi$ production at the Tevatron \cite{cdf}. 
This has led to a major revision of the theory of quarkonium formation. 
In the colour-singlet model, the relative velocity, $v$ 
between the charm quarks is neglected. However, for charmonia
$v^2 \sim 0.3$, and is not negligible. It is possible to include the 
effects coming from higher orders in a perturbation series in $v$ 
in non-relativistic QCD (NRQCD) \cite{caswell, bbl}. 

In this picture, the physical charmonium state can be expanded 
in terms of its Fock-components in a perturbation series in $v$, 
and the $c \bar c$ states appear in either colour-singlet or 
colour-octet configurations in this series. The 
colour-octet $c \bar c$ state makes a transition into a physical
state by the emission of one or more soft gluons, 
dominantly $via$ $E1$ or $M1$ transitions with
higher multipoles being suppressed by powers of $v$. 
The cross-section for the production of a meson $H$ 
then takes on the following factorised form:
\begin{equation}
   \sigma(H)\;=\;\sum_{n=\{\alpha,S,L,J\}} {F_n\over m^{d_n-4}}
       \langle{\cal O}^H_\alpha({}^{2S+1}L_J)\rangle
\label{e1}
\end{equation}
where $F_n$'s are the short-distance coefficients and ${\cal O}_n$ are local
4-fermion operators, of naive dimension $d_n$, describing the long-distance
physics. 

In fact, the importance of the colour-octet components was first noted
\cite{bbl2} in the case of $P$-wave charmonium decays, and even in
the production case the importance of these components was first
seen \cite{jpsi} in the production of $P$-state charmonia at the
Tevatron. The surprise was that even for the production of $S$-states
such as the $J/\psi$ or $\psi'$, where the colour singlet components give
the leading contribution in $v$, the inclusion of sub-leading octet states
was seen to be necessary for phenomenological reasons \cite{brfl}. 
While the inclusion of the colour-octet components seem to be
necessiated by the Tevatron charmonium data, the normalisation of
these data cannot be predicted because the the long-distance matrix 
elements are not calculable. The data allow a linear combination of 
octet matrix-elements to be fixed \cite{cgmp,cho}, and much effort
has made recently to understand the implications of these colour-octet
channels for $J/\psi$ production in other experiments \cite{allref}.

One important feature of the NRQCD Lagrangian is that it shows
an approximate heavy-quark symmetry, which is valid to $O(v^2) \sim 0.3$.
The implication of this symmetry is that the nonperturbative parameters
have a weak dependence on the magnetic quantum number. We
show that the production cross-section of the ${}^1P_1$ charmonium 
state, $h_c$, can be $predicted$ in NRQCD, because of the heavy-quark
symmetry. The Tevatron data on $\chi_c$ production fixes \cite{cho} the 
colour-octet matrix element which specifies the transition of a 
${}^3S_1$ octet state into a ${}^3P_J$ state. We would expect from 
heavy-quark spin symmetry of the NRQCD Lagrangian that the 
matrix-element for ${}^1S_0^{\lbrack 8 \rbrack} \rightarrow h_c$ should be
of the same order as that for ${}^3S_1^{\lbrack 8 \rbrack} \rightarrow 
{}^3P_1$. The production of the ${}^1P_1$ charmonium state, $h_c$,
is, therefore, a crucial prediction of NRQCD.

The production of the $h_c$ is interesting in its own right~: 
charmonium spectroscopy \cite{onep} predicts this state to exist at 
the centre-of-gravity of the $\chi_c ({}^3P_J)$ states. While the E760 
collaboration at the Fermilab has reported \cite{e760} the first 
observation of this resonance its existence needs further confirmation. 
In the following, we report on the calculations for $h_c$ production 
at the Tevatron presented in Ref.~\cite{mine}.

Being a $P$-state, the leading colour-singlet contribution is
already at $O(v^2)$, and at the same order we have the octet
production of the ${}^1P_1$ state through an intermediate 
${}^1S_0$ state. The octet channel is, therefore, expected to
be very important for the production of this state. Once produced, the
$h_c$ can be detected by its decay into a $J/\psi + \pi^0$.
The subprocesses that we are interested in are the following:
\begin{eqnarray}
g + g &\rightarrow {}^1P_1^{\lbrack 1 \rbrack} + g, \nonumber \\
g + g &\rightarrow {}^1S_0^{\lbrack 8 \rbrack} + g, \nonumber \\
q(\bar q)+ g &\rightarrow {}^1S_0^{\lbrack 8 \rbrack } + 
                  q(\bar q), \nonumber \\
q + \bar q &\rightarrow {}^1S_0^{\lbrack 8 \rbrack } + g.
      \label{e2}
\end{eqnarray}
The ${}^1S_0^{\lbrack 8 \rbrack} \rightarrow h_c$ is mediated by
a gluon emission in a $E1$ transition. 

The cross-section $d\sigma / dp_T$ for $h_c$ production at the 
Tevatron energy ($\sqrt{s}= 1.8$~TeV) presented in Ref.~\cite{mine} 
are shown in Fig.~1. For 20~pb${}^{-1}$ total luminosity, for $p_T$
integrated between 5 and 20~GeV we expect of the order of 650
events in the $J/\psi + \pi$ channel. Of these, the contribution 
from the colour-singlet channel is a little more than 40, while the 
octet channel gives more than 600 events. The colour-octet dominance 
is more pronounced at large-$p_T$. Recent results on $J/\psi$ production
from CDF are based on a total luminosity of 110~pb${}^{-1}$. For
this sample, more than 3000 events can be expected to come
from the decay of the $h_c$ into a $J/\psi$ and a $\pi$. With this
large event rate, the $h_c$ should certainly be observable if the
$pi^0$ coming from its decay can be reconstructed efficiently. 

In conclusion, we find that a reasonably large rate for the production
of the ${}^1P_1$ is expected at the Tevatron, with a dominant contribution
from the colour-octet production channel. Heavy-quark symmetry relations
allow us to infer the size of the non-perturbative matrix-elements and 
to predict the rate for $h_c$ production and its subsequent
decay into a $J/\psi$ and a $\pi$. By looking at $J/\psi$ events associated
with a soft pion, it should be possible to pin down the elusive ${}^1P_1$
resonance at the Tevatron, given the large number of $J/\psi$ events
already available. We believe that this is a firm prediction of the
NRQCD framework and may be a useful way of distinguishing this from
other models of quarkonium formation.

\begin{figure}[h]
\vskip 6.2in\relax\noindent
          \relax{\includegraphics{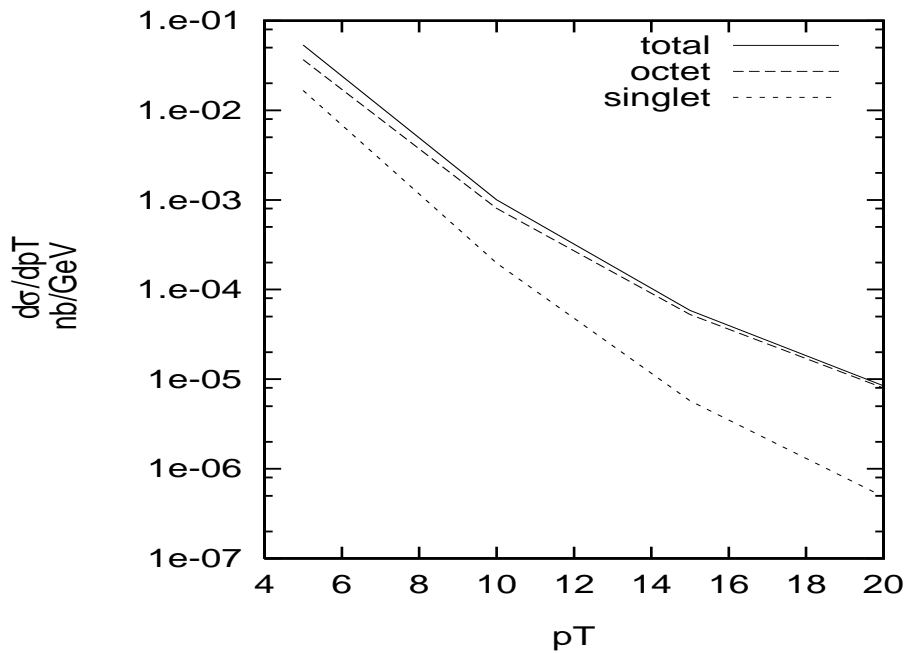}}  
\vspace{-40ex}
\caption{$d\sigma/dp_T$ (in nb/GeV) for $h_c$ production (after folding in
with Br($h_c \rightarrow J/\psi+\pi$)=0.5) at
1.8~TeV c.m. energy with $-0.6 \le y \le 0.6$. }
\end{figure}

\end{document}